# HIGH POWER PULSES EXTRACTED FROM THE PEREGRINE ROGUE WAVE

GUANGYE YANG,[1,2] LU LI,[1,*] SUOTANG JIA,[3] DUMITRU MIHALACHE [4,5]

[1] Shanxi University, Institute of Theoretical Physics, Taiyuan, 030006, China
[2] Shanxi Medical University, Department of Physics, Taiyuan, 030001, China
[3] Shanxi University, College of Physics and Electronics Engineering, Taiyuan, 030006, China
[4] Academy of Romanian Scientists, 54 Splaiul Independentei, RO-050094 Bucharest, Romania
[5] "Horia Hulubei" National Institute of Physics and Nuclear Engineering, Magurele-Bucharest, Romania
*E-mail: llz@sxu.edu.cn

*Abstract.* We address various initial excitations of the Peregrine rogue wave and establish a robust transmission scheme of high power pulses extracted from the Peregrine rogue wave in a standard telecommunications fiber. The results show that the Peregrine rogue wave can be excited by using a weak pulse atop a continuous wave background and that the high power pulses extracted from the Peregrine rogue wave exhibit the typical characteristics of breathing solitons. The influence of higher-order effects, such as the third-order dispersion, the self-steepening and the Raman effect, on the propagation of the pulse extracted from the peak position and the interaction between neighboring high power pulses induced by initial perturbations are also investigated.

*Key words*: high power pulses, Peregrine rogue wave, nonlinear Schrödinger equation.

## 1. INTRODUCTION

The nonlinear Schrödinger (NLS) equation is an important model in optical fiber transmission. Recently, based on this model, the rogue wave phenomenon has been extensively studied, including rational solutions, pulse splitting induced by modulation instability and wave turbulence [1–3]. A fundamental analytical solution on the rogue wave is the Peregrine solution, which was first presented by Peregrine [4]. The Peregrine solution (here it is also called the Peregrine rogue wave) is a localized solution in both time and space, and is a limiting case of Kuznetsov-Ma soliton [5–7] and Akhmediev breather [8, 9]. The former case exhibits a periodization process of the bright soliton and the latter case features a localized process of the continuous wave (CW) [10, 11]. Other recent theoretical works investigate the persistence of rogue waves [12, 13] in the integrable Sasa-Satsuma equation [14, 15] and the general rogue waves in the Davey-Stewartson I equation [16]. The rogue wave phenomenon has been intensively studied, including nonautonomous "rogons" in the inhomogeneous nonlinear Schrödinger equation with variable coefficients [17], N-order bright and dark rogue waves in resonant erbium-doped fiber systems [18], rogue-wave solutions of a three-component coupled nonlinear Schrödinger equation [19], and rogue waves of the Hirota and the Maxwell-Bloch equations [20]. It should be noted that the majority of reported results in the area of rogue waves are theoretically described by the NLS equation and its modifications that are necessary in the subpicosecond region, where we must consider higher-order effects, such as third-order dispersion, self-steepening and self-frequency shift [12]. Such studies might be extended to other relevant dynamical models describing the formation of ultrashort temporal solitons in the femtosecond domain; see a recent review of several non-NLS-type models for ultrashort solitons, beyond the slowly varying envelope approximation [21]. Relevant experiments on the Peregrine soliton generation in telecommunications fibers and the observation of Kuznetsov-Ma soliton dynamics in optical fibers have been recently reported [22-24].

Akhmediev breather is a family of exact periodic solution in transverse dimension for NLS equation [8], which can describe the evolution of initially continuous waves and can be excited by a weak periodic modulation [25]. Based on the Akhmediev breather, the continuous wave supercontinuum generation from modulation instability and Fermi-Pasta-Ulam (FPU) recurrence have been examined [26-28]. It should be pointed out that the initial excitation of the Akhmediev breather is mainly restricted by perturbed cosine-type function, i. e., a weak periodic modulation. Also, possibility of an Akhmediev breather decaying into solitons has been studied recently [29]. In contrast to the Akhmediev breather, Kuznetsov-Ma soliton is a localized solution in transverse dimension for NLS equation, and can be excited by a small localized (single peak) perturbation pulse of CW background [11] or by strongly modulated continuous wave [24]. Motivated by that, in this paper, we will discuss the various initial excitations with the single peak modulation for the Peregrine rogue wave, and establish a robust transmission scheme induced by the Peregrine rogue wave in optical fibers.

## 2. EXCITATION OF THE PEREGRINE ROGUE WAVE

In the picosecond regime, the propagation of optical pulse inside single-mode fibers can be governed by the NLS equation as follows [30]

$$\mathrm{i}\frac{\partial A}{\partial z} - \frac{\beta_2}{2}\frac{\partial^2 A}{\partial \tau^2} + \gamma |A|^2 A = -\frac{\mathrm{i}}{2}\alpha A, \quad (1)$$

where $A(z,\tau)$ is the slowly varying amplitude of the pulse envelope, $z$ is the propagation distance and $\tau$ is the time in a frame of reference moving with the pulse at the group velocity $v_g$ ($\tau = t - z/v_g$), and $\alpha$ is related to the loss/gain ($\alpha > 0$ for the loss and $\alpha < 0$ for the gain). In the absence of the loss/gain, i. e., $\alpha = 0$, Eq. (1) has a rational fraction solution as follows [22, 23]

$$A(z,\tau) = \sqrt{P_0}\, e^{iZ}\left[1 + R(z,\tau)\right], \qquad (2)$$

where $R(z,\tau) = e^{i\pi}(4 + i8Z)/(1 + 4Z^2 + 4T^2)$, and $T = (\tau - \tau_0)/T_0$ and $Z = (z - z_0)/L_{NL}$ with $T_0 = [|\beta_2|/(\gamma P_0)]^{1/2}$, $L_{NL} = (\gamma P_0)^{-1}$, where $z_0$ and $\tau_0$ are arbitrary real constants. The solution (2) was first derived from the Kuznetsov-Ma breather by Peregrine [4, 5, 7] and is called the Peregrine solution or the Peregrine rogue wave. From Eq. (2), one can see that the Peregrine solution (2) is a superposition of a CW solution for Eq. (1) with $\alpha = 0$ and a rational fraction function $R(z,\tau)$, and reaches the peak value at $z = z_0$ and $\tau = \tau_0$. This implies that the Peregrine rogue wave can be excited by an initial perturbation $R(0,\tau)$ on the CW background. To study the initial excitation of the Peregrine rogue wave, here we consider a standard silica SMF-28 fiber at 1550 nm with the group velocity dispersion $\beta_2 = -21.4$ ps²/km, the nonlinear parameter $\gamma = 1.2$ W$^{-1}$km$^{-1}$ and the loss $\alpha_{dB} = 0.19$ dB/km with $\alpha_{dB} = 4.343\alpha$ [2, 23].

From the expression of the rational function $R(z,\tau)$, one can see that when $z = 0$, its initial FWHM, peak intensity and phase are of the form $W_0 = [(\sqrt{2}-1)(1 + 4z_0^2/L_{NL}^2)]^{1/2}T_0$, $\varepsilon = 4/(1 + 4z_0^2/L_{NL}^2)^{1/2}$, and $\theta_0 = \pi - \arctan(2z_0/L_{NL})$, respectively, in which the initial intensity and the initial phase are dimensionless. To compare the initial excitation characteristics of the Peregrine rogue wave, we also take the Gaussian-type pulse and the hyperbolic secant-type pulse as the initial perturbations on the CW to excite the Peregrine rogue wave, which are of the form $R_G(\tau) = \varepsilon \exp[-(\tau - \tau_0)^2/2W_G^2]e^{i\theta_0}$ and $R_S(\tau) = \varepsilon \text{sech}[(\tau - \tau_0)/W_S]e^{i\theta_0}$, respectively, where $W_G = W_0/(2\sqrt{\ln 2})$ and $W_S = W_0/\ln(2\sqrt{2}+3)$, in which the initial FWHM, peak intensity and phase are kept.

Figure 1 presents the evolution plots of the three kinds of initial conditions. From Fig. 1a, one can see that the localized rational function perturbation $R(0,\tau)$ on CW background can excite an ideal Peregrine rogue wave in the absence of loss ($\alpha = 0$). When considering the system's loss, the Peregrine rogue wave can be still excited, but after forming it, the Peregrine rogue wave is split, as shown in Fig. 1b. Furthermore, even if the initial perturbations deviate from the rational function, such as replacing the localized rational function $R(0,\tau)$ as the Gaussian-type perturbation $R_G(\tau)$ and the hyperbolic secant-type perturbation $R_S(\tau)$ with the same initial FWHM, peak intensity and phase, the Peregrine rogue wave can be also formed, and similarly after forming it, the Peregrine rogue wave is split, as shown in Figs. 1c and 1d. Here, for comparison, Figs. 1e and 1f present the intensity profiles at the initial position and the peak position. From them, one can see that their profiles are almost the same. Comparing the results in Ref. [16], in which the Peregrine rogue wave is excited by a cosine-type function perturbation, here it is shown that the Peregrine rogue wave can be also excited by a small localized (single peak) perturbation, i. e., a weak pulse atop a continuous wave background.

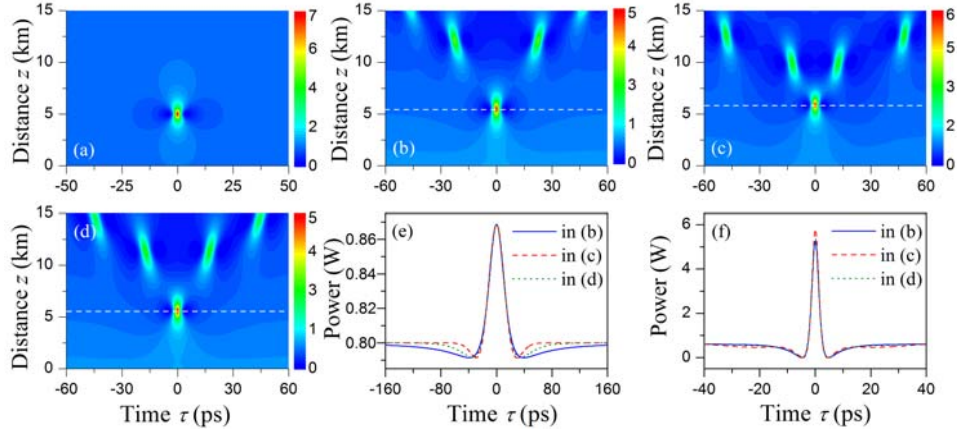

Fig. 1 – Evolution plots of the different initial conditions; a) and b) for the initial condition $R(0, \tau)$ without and with loss ($\alpha = 0$ and $\alpha_{db} = 0.19$ dB/km), respectively; c) for the initial condition $R_G(\tau)$ with $\alpha \neq 0$; d) for the initial condition $R_S(\tau)$ with $\alpha \neq 0$; e) intensity profile at the initial position; f) distributions of the intensity profile at peak position $z_p = 5.475$ km in (b), $z_p = 5.879$ km in (c) and $z_p = 5.596$ km in (d), respectively. Here the parameters are $P_0 = 0.8$ W, $\tau_0 = 0$, and $z_0 = 5$. The color bars inserted in this figure and in the following ones show the corresponding powers measured in W.

The above results reveal that the Peregrine rogue wave can be conveniently excited by a small localized (single peak) perturbation of the CW, but the peak positions are different for the different initial excitations. Indeed, for our choice of the parameters, the peak positions are, respectively, $z_p = 5$ km, $z_p = 5.38$ km, and $z_p = 5.113$ km in the absence of loss. Considering the loss, they are $z_p = 5.475$ km, $z_p = 5.879$ km and $z_p = 5.596$ km, respectively, as shown in Fig. 1. Furthermore, Fig. 2 shows the dependence of the peak position $z_p$ on the initial power $P_0$ and $z_0$ for different initial excitations. From it one can see that the peak position $z_p$ is decreasing as a function of the initial power $P_0$ and is increasing as a function of $z_0$ for the three initial excitations. Also, the system's loss delayed the occurrence of the peak position. Thus it is easy to determine the peak position for a given initial condition based on the curves in Fig. 2.

Finally, it should be pointed out that the results shown in Figs. 1 and 2 are valid for the initial perturbation pulses with the FWHM, peak intensity and phase derived from Eq. (2). Furthermore, we also performed numerical simulations for the generation of the Peregrine rogue wave when the FWHM, peak intensity and phase of the initial perturbation pulses deviate from the exact analytical values derived from Eq. (2). The numerical simulations show that similar results hold, but the corresponding peak's positions are different.

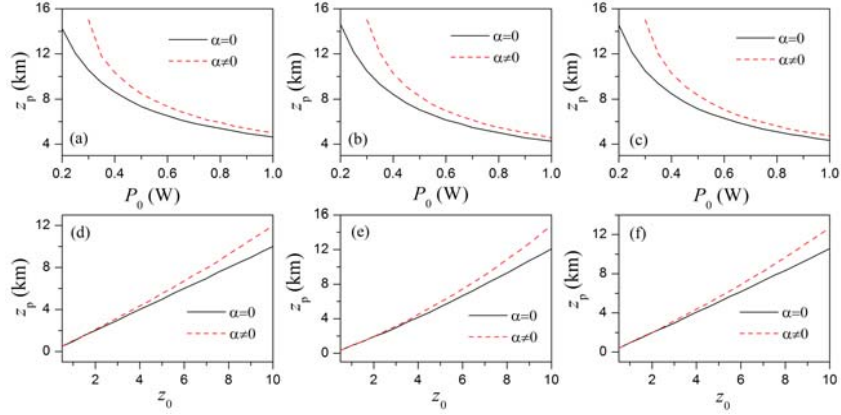

Fig. 2 – Dependence of the peak position $z_p$ on the initial power $P_0$ and $z_0$ for: a) and d) the rational function initial excitation; b) and e) the Gaussian-type initial excitation; c) and f) the hyperbolic secant-type initial excitation, respectively. Here $z_0 = 5$ and $\tau_0 = 0$ in (a), (b) and (c); $P_0 = 0.8$W and $\tau_0 = 0$ in (d), (e) and (f).

## 3. HIGH POWER PULSES EXTRACTED FROM THE PEREGRINE ROGUE WAVE AND THEIR INTERACTRIONS

It should be emphasized that the Peregrine rogue wave is maximally compressed at the peak position, thus a maximally compressed pulse with high power can be generated by a small localized (single peak) perturbation of the CW, although the peak positions are different for the different initial excitations, as shown in Fig. 2. However, as applications, the maximally compressed pulse occurred at the peak position can not directly travel with preserving shape, and so can not exhibit a robust transmission scheme in the optical fiber. This is because of the presence of the background wave $q_B(z) = \sqrt{P_0}$ at the peak position from the expression (2), which leads to FPU-like growth-return evolution and the pulse splitting due to the modulation instability. To apply the Peregrine rogue wave to generate high power pulses with preserving shape, one must eliminate the background wave $q_B(z) = \sqrt{P_0}$ from the Peregrine rogue wave at the peak position. According to this idea, we performed propagation simulations for different initial excitations by subtracting the background wave of the form $q_B(z) = \sqrt{P_0}$ from the maximally compressed pulse at the peak position, and the results are summarized in Fig. 3. Note that subtracting the background wave of another form $q'_B(z) = \sqrt{P_0}\,\mathrm{e}^{\mathrm{i}(z_p - z_0)/L_{NL}}$ from the maximally compressed pulse at the peak position we obtain similar results to those shown in Fig. 3. This means that a small change of the phase of the eliminated background wave do not strongly influence the propagation features of the high power pulse extracted from the peak position. Thus an initial perturbation on CW excited a high power pulse on the background wave $q_B(z)$ at the peak position $z_p$, and then after eliminating the background wave, the pulse extracted from the peak position is abruptly amplified and exhibits the oscillatory propagation feature, forming a *breathing soliton*. Furthermore, we also investigated the propagation of pulses extracted from the position $z = z_1 < z_p$, which deviates less from the peak position $z_p$. In this case the background wave in the position $z_1$ is of the form $q_B(z) = \sqrt{P_0}\,e^{i(z_1-z_0)/L_{NL}}$, which includes a phase factor. So we should eliminate a background wave with the phase factor at the position $z_1$, thus the evolution of the pulse extracted from $z_1$ exhibits similar results to those shown in Fig. 3. The only difference with the results plotted in Fig. 3 is that the power of the obtained breathing soliton is diminished, its oscillatory period becomes larger and the average width becomes wider. Also, when considering the fiber loss, the evolution characteristics of the breathing solitons do not change except for an attenuation of the pulse power.

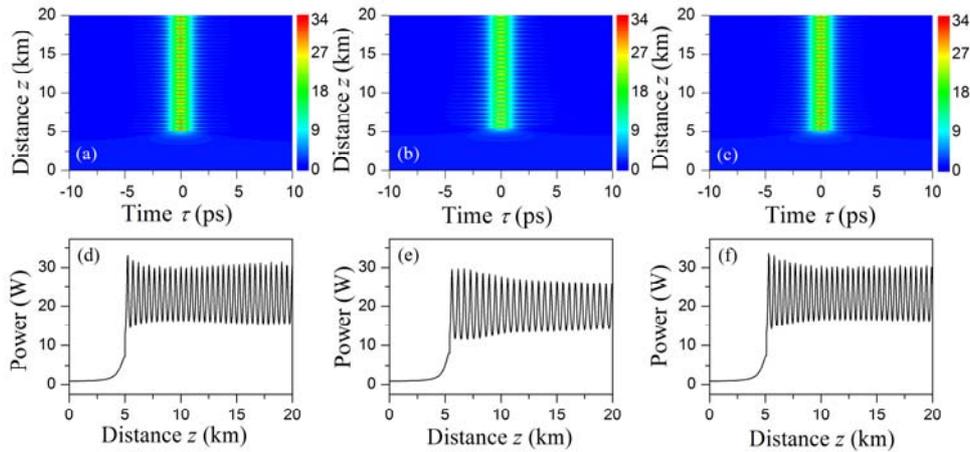

Fig. 3 – Propagation evolution plots of the high power pulse extracted from the peak position and the corresponding power evolution plots for: a) and d) the rational function initial excitation; b) and e) the Gaussian-type initial excitation; c) and f) the hyperbolic secant-type initial excitation, respectively. Here $P_0 = 0.8$ W, $\tau_0 = 0$, and $z_0 = 5$.

We also studied the influence of the higher-order terms in the NLS equation (1), such as the third-order dispersion, the self-steepening and the Raman effect [11, 30], on the pulse extracted from the peak position. The obtained results are similar to those shown in Fig. 3 except for some time shift, as shown in Fig. 4. In Figs. 4a and 4b we only consider the influence of the third-order dispersion, whereas in Figs. 4c and 4d we include the third-order dispersion, the self-steepening and the Raman effect.

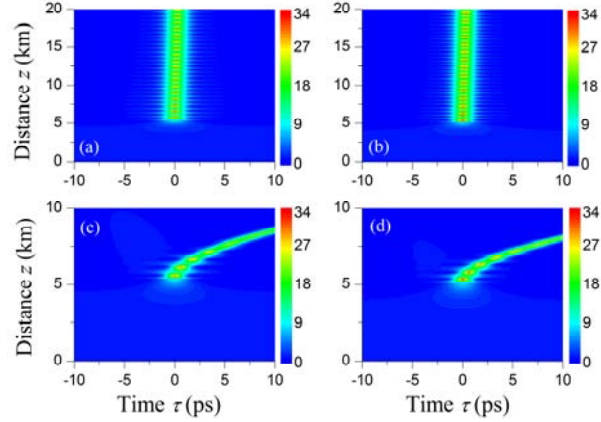

Fig. 4 – Propagation evolution plots of the high power pulse extracted from the peak position for: a) and c) the Gaussian-type initial excitation; b) and d) the hyperbolic secant-type initial excitation, respectively, when considering the higher-order effects. Here (a) and (b) only include the third-order dispersion, which is $\beta_3 = 0.12$ ps$^3$/km; c) and d) include the third-order dispersion, the self-steepening and the Raman effect, where $\beta_3 = 0.12$ ps$^3$/km and the Raman time constant $T_R = 5$ ps, respectively. Other parameters are $P_0 = 0.8$ W, $\tau_0 = 0$, and $z_0 = 5$.

Finally, we investigate the interaction between neighboring high power pulses extracted from the Peregrine rogue wave. Here, we consider the initial condition in the form

$$A(0,\tau) = \sqrt{P_0}\, e^{-iZ_0} \left[ 1 + R_{G/S}\left(\tau + T_P/2\right) + R_{G/S}\left(\tau - T_P/2\right) e^{i\varphi_0} \right], \tag{3}$$

where $T_P$ is the initial separation between neighboring pulses, $R_{G/S}$ represents either the Gaussian-type perturbation or the hyperbolic secant-type perturbation, $\varphi_0$ is the phase difference between neighboring pulses, and $Z_0 = z_0/L_{NL}$. Figure 5 plots the evolutions of the neighboring pulses for both types of initial excitations, i. e., for the hyperbolic secant-type and the Gaussian-type initial excitations, under in-phase condition, i.e., $\varphi_0 = 0$. It is found that when the initial separation is relatively small, for the hyperbolic secant-type initial perturbation such as $T_P = 28$ ps as shown in Fig. 5a, the two neighboring high power pulses attract each other, then collide, and finally repulse each other, a result which resembles a typical elastic collision, while for the Gaussian-type initial perturbation such as $T_P = 32$ ps in Fig. 5e the two pulses have some irregular propagation dynamics due to strong interaction. When the initial separation $T_P$ is further increased, the interaction will be gradually weakened, as shown in Figs. 5b and 5f with the initial separation $T_P = 40$ ps. When the initial separation reaches $T_P = 60$ ps, the corresponding separation between the two pulses extracted from the peak position remains practically unchanged during propagation, as shown in Figs. 5c and 5g. It should be pointed out that during propagation the pulses still exhibit the intrinsic oscillatory behavior. Furthermore, we performed the propagation of 1-byte pulse train (*i.e.*, eight pulses) with the initial separation $T_P = 60$ ps for both the hyperbolic secant-type and the Gaussian-type initial excitation, as shown in Figs. 5d and 5h, respectively. Thus from Figs. 5d and 5h one can clearly see that there is practically no interaction during propagation. The propagation of such pulse trains without interaction provides an effective means for using such high power pulses as information carriers in ultrahigh capacity optical communication systems so that the maximum possible transmission rate can attain about 2GB/s.

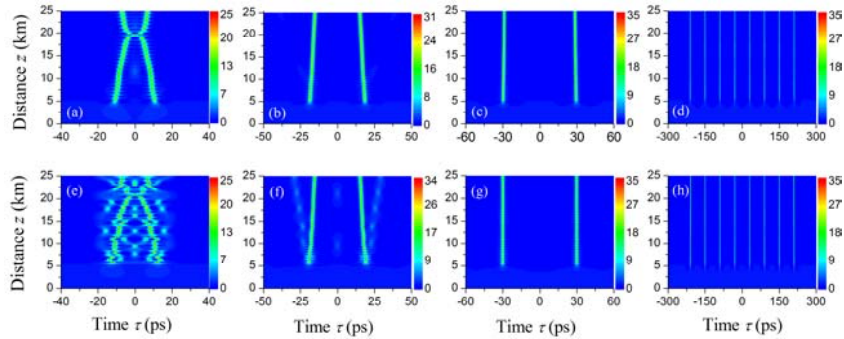

Fig. 5 – Contour plots of the interaction between in-phase neighboring high power pulses excited by: a)–d) the hyperbolic secant-type initial perturbation; e–h the Gaussian-type initial perturbation for different separations, respectively. Here a) $T_P$ =28 ps, e) $T_P$ =32 ps; b) and f) $T_P$ =40; c,d,g,h $T_P$ = 60 ps. The other parameters are the same as in Fig. 3.

In the case of interactions between out-of-phase neighboring pulses, there are some interesting features as shown in Fig. 6. Comparing with Fig. 5, one can see that in the case of the hyperbolic secant-type initial perturbation, one of the neighboring pulses is weakened or even eliminated, while in the case of the Gaussian-type initial perturbation, the two pulses mainly exhibit a repulsive interaction.

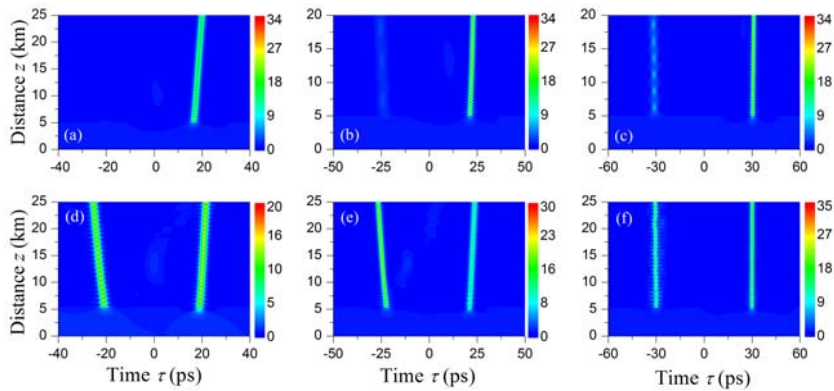

Fig. 6 – Contour plots of the interaction between out-of-phase neighboring high power pulses excited by: a)–c) the hyperbolic secant-type initial perturbation; d)–f) the Gaussian-type initial perturbation with out-of-phase for different separations, respectively. Here: a) $T_P$ = 28 ps; d) $T_P$ = 32 ps; b) and e) $T_P$ = 40; c) and f) $T_P$ = 60 ps. The other parameters are the same as in Fig. 3.

## 4. CONCLUSIONS

In summary, based on the NLS model, we have studied various excitations of the Peregrine rogue wave, and have established a robust transmission scheme of high power pulses extracted from the Peregrine rogue wave in the fiber system. The results show that these robust pulses exhibit oscillatory (breathing) propagation feature. Finally, the interaction between both in-phase and out-of-phase neighboring pulses under initial perturbations and the influence on propagation of the third-order dispersion, the self-steepening and the Raman effect have been also investigated. Our results may provide a convenient way to generate robust high power pulses extracted from the Peregrine rogue wave.

*Acknowledgments.* This research was supported by the National Natural Science Foundation of China under Grant No. 61078079 and the Shanxi Scholarship Council of China under Grant No. 2011-010. The work of D.M. was supported in part by Romanian Ministry of Education and Research, CNCS-UEFISCDI, project number PN-II-ID-PCE-2011-3-0083.